# High sensitivity multi-axes rotation sensing using large momentum transfer point source atom interferometry

Jinyang Li[1], Gregório R. M. da Silva[1], Wayne C. Huang[1], Mohamed Fouda[3], Timothy Kovachy[1] and Selim M. Shahriar[1,2]

[1]Department of Physics and Astronomy, Northwestern University, Evanston, IL 60208, USA
[2]Department of ECE, Northwestern University, Evanston, IL 60208, USA
[3]Digital Optics Technologies, Rolling Meadows, IL 60008, USA
Email: shahriar@northwestern.edu

**Abstract**

A point source interferometer (PSI) is a device where atoms are split and recombined by applying a temporal sequence of Raman pulses during the expansion of a cloud of cold atoms behaving approximately as a point source. The PSI can work as a sensitive multi-axes gyroscope that can automatically filter out the signal from accelerations. The phase shift arising from rotations is proportional to the momentum transferred to each atom from the Raman pulses. Therefore, by increasing the momentum transfer, it should be possibly to enhance the sensitivity of the PSI. Here, we investigate the degree of enhancement in sensitivity that could be achieved by augmenting the PSI with large momentum transfer (LMT) employing a sequence of many Raman pulses with alternating directions. Contrary to typical approaches used for describing a PSI, we employ a model under which the motion of the center of mass of each atom is described quantum mechanically. We show how increasing Doppler shifts lead to imperfections, thereby limiting the visibility of the signal fringes, and identify ways to suppress this effect by increasing the effective, two-photon Rabi frequencies of the Raman pulses. Taking into account the effect of spontaneous emission, we show that, for a given value of the one-photon Rabi frequency, there is an optimum value for the number of pulses employed, beyond which the net enhancement in sensitivity begins to decrease. For a one-photon Rabi frequency of 200 MHz, for example, the peak value of the factor of enhancement in sensitivity is ~39, for a momentum transfer that is ~69 times as large as that for a conventional PSI. We also find that this peak value scales as the one-photon Rabi frequency to the power of 4/5.

# 1. Introduction

Atom interferometry offers the potential to deliver high-performance, compact, and robust gyroscopes that are suitable for inertial navigation applications. Critical requirements for such an atomic gyroscope include a high sensitivity to rotations, and the ability to distinguish between signals arising from rotations and accelerations. Here, we describe a multi-axes gyroscope based on the combination of point source interferometry (PSI)[1,2,3,4] and large momentum transfer (LMT) beam splitters[5,6] which is well-suited to meet these requirements. In a PSI, Raman pulses are applied during the expansion of a point source of atoms. The pulses are a pair of counter-propagating laser beams that drive two-photon Raman transitions, serving as the beam splitters and mirrors for a Mach-Zehnder light-pulse atom interferometer[7,8], as shown in Fig. 1. The interferometer phase response to rotation scales linearly with the velocity difference of atoms in the two arms, while the response to acceleration is independent of the atomic velocity. Because of this difference, the signal in a PSI allows rotation and acceleration to be distinguished. The PSI can also determine both components of the rotation vector that are orthogonal to the laser pulses, thus realizing a multi-axes gyroscope. The LMT beam splitters we consider involve the use of tailored laser pulse sequences to increase the momentum splitting, and therefore the velocity difference, between the two arms of the interferometer. Via the Sagnac effect, the rotation sensitivity of a gyroscope is proportional to the area enclosed by an interferometer. The enclosed area is proportional to the velocity difference induced by the beam splitter; as such, the rotation sensitivity scales linearly with the momentum transferred by the laser pulses during the beam splitting process.

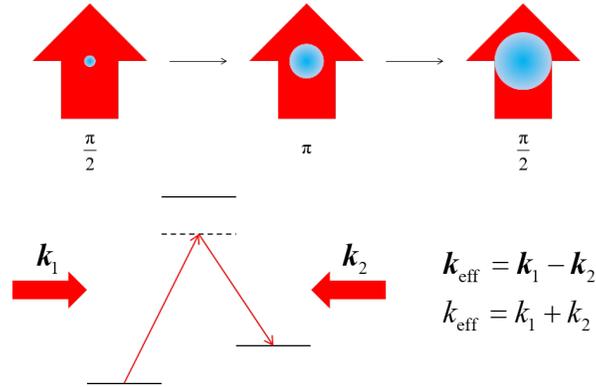

Fig. 1 Schematic illustration of the basic process underlying the conventional PSI. The blue circle is the atom cloud and the red arrows are Raman pulses. A temporal sequence of Raman pulses is applied during the expansion of the atom cloud (top), with each Raman pulse being a pair of counter-propagating light beams that drives a two-photon transition (bottom).

The conventional model of a PSI makes the approximation that each atom has a well-defined velocity as well as a well-defined position. However, this model is inadequate for describing the behavior of a PSI accurately, for several reasons. The first is that the wave packets of cold atoms are widespread, and thus do not have trajectories that enclose a well-defined area. The second is that atoms are in superpositions of many momentum eigenstates, with each of them seeing a different light frequency. As such, it necessary to treat the center of mass motion of each atom as a wave packet in order to determine the nature of the signal for an LMT-PSI, and its sensitivity compared with that of a regular PSI.

## 2. Conventional model

As noted above, the conventional model of a PSI makes the approximation that each atom has a well-defined velocity as well as a well-defined position. Therefore, the atoms follow definitive trajectories that enclose an area. Specifically, the enclosed area is $A = (r/2) \times (\hbar k_t T/m)$, where $\hbar k_t$ is the momentum transferred to an atom from the initial light pulse, $r$ is the displacement of the atoms, $T$ is half of the total time elapsed, from splitting to recombination, and $m$ is the mass of each atom[9]. The Sagnac phase shift is proportional to the enclosed area according to the expression $\phi = 2\omega_C \mathbf{\Omega} \cdot \mathbf{A}/c^2$, where $\omega_C = mc^2/\hbar$ is the Compton frequency of each atom[10], and $\mathbf{\Omega}$ is the angular velocity of the rotation. It then follows that the phase shift can be expressed as have $\phi = (\mathbf{k}_t \times \mathbf{\Omega} T) \cdot \mathbf{r} \equiv \mathbf{k}_\Omega \cdot \mathbf{r}$. The measured signal is the spatial distribution of the ground state population, given by the expectation value of the projection operator $P_g(\mathbf{r}) \equiv |g,\mathbf{r}\rangle\langle g,\mathbf{r}|$. As such, the signal can be expressed[9] as $\langle P_g(\mathbf{r}) \rangle = (1 + \cos \mathbf{k}_\Omega \cdot \mathbf{r})/2$, which is a pattern of spatial fringes dictated by the wave number $\mathbf{k}_\Omega$. With this model, it seems obvious that by increasing $\mathbf{k}_t$, we can increase $\mathbf{k}_\Omega$, thus reducing the fringe spacing, and thereby increasing the sensitivity of the PSI. However, it is not obvious whether this model is valid for a PSI with an arbitrary $\mathbf{k}_t$, because the wave packets of the atoms are widespread, making the enclosed area ill-defined. In addition, this conventional model cannot predict loss of the signal contrast due to the quantum nature of the center of mass motion of the atoms. Therefore, it is necessary to build a model that treats the center of mass motion of each atom quantum mechanically, represented as a wave packet[11].

## 3. Quantum model

The quantum state of an atom consists of its internal state and the state of its center of mass. Each atom in a PSI starts with its internal state being the ground state $|g\rangle$. We first consider the case where the center of mass of the atom is initially in a momentum eigenstate $|\mathbf{k}\rangle$. The evolution of the state of the center of mass is illustrated in Fig. 2. The first atomic beam splitter propagating in the $z-$direction splits the atom into a superposition of the state $|\mathbf{k} - k^*\hat{z}\rangle$ and the state $|\mathbf{k} + (k_t - k^*)\hat{z}\rangle$, where the value of $k^*$ depends on the details of the technique employed for LMT. In a certain picture, the energy of the state $|\mathbf{k} - k^*\hat{z}\rangle$ and the state $|\mathbf{k} + (k_t - k^*)\hat{z}\rangle$ can be the same, and can be defined to be zero. For simplicity, we work in such a picture. The atomic mirror pulse is applied at a time $T$ after the first atomic beam splitter. Here, the effect of rotation comes in. Due to rotation perpendicular to the $z-$direction, the atomic mirror pulses are no longer in the original $z-$direction, but at an angle $\Omega T$ with respect to it. We assume that the angular velocity of the rotation is in the $x-$direction, without loss of generality. For $\Omega T \ll 1$, the atomic mirror pulses turn each atom into a superposition of the state $|\mathbf{k} - k^*\hat{z} + k_t \Omega T \hat{y}\rangle \equiv |+\rangle$ and the state $|\mathbf{k} + (k_t - k^*)\hat{z} - k_t \Omega T \hat{y}\rangle \equiv |-\rangle$. The last atomic beam splitter will combine these two states approximately back to the position $\mathbf{k}$ in the momentum space. The Sagnac phase shift can be viewed as arising from the energy difference between states $|+\rangle$ and $|-\rangle$. To see this, note first that the energies of these two states are no longer zero, but $\hbar^2\left[\left(k_t \Omega T\right)^2 \pm 2k_t \Omega T \boldsymbol{k}\cdot\hat{\boldsymbol{y}}\right]/2m$. Therefore, these two states oscillate at frequencies that have a difference of $2\hbar k_t \Omega T \boldsymbol{k}\cdot\hat{\boldsymbol{y}}/m$. The Sagnac phase shift is the product of this frequency difference and $T$, the duration for the second half of the interferometry process: $\phi = 2\hbar k_t \Omega T^2 \boldsymbol{k}\cdot\hat{\boldsymbol{y}}/m$. It can also be written as $\phi = 2\hbar\left(\boldsymbol{k}_t \times \boldsymbol{\Omega} T^2 / m\right)\cdot\boldsymbol{k} \equiv \boldsymbol{r}_\Omega \cdot \boldsymbol{k}$, where $\boldsymbol{k}_t \equiv k_t\hat{\boldsymbol{z}}$. Note that $r_\Omega = 2\hbar k_\Omega T / m$, and while it has the dimension of distance, it does not represent the spatial coordinate of the center of mass of the atom. It can be shown that this expression of the phase shift is equivalent to the phase shift under the conventional description if we assign to a momentum eigenstate $|\boldsymbol{k}'\rangle$ a localized state with a velocity of $\hbar\boldsymbol{k}'/m$, determine the vectorial area $\boldsymbol{A}$ enclosed by the resulting trajectories, and use the expression $\phi = 2m\boldsymbol{\Omega}\cdot\boldsymbol{A}/\hbar$.

Due to the rotation induced phase shift, the population of the ground state will be $\langle P_g(\boldsymbol{k})\rangle = \cos^2\left(\boldsymbol{r}_\Omega \cdot \boldsymbol{k}/2\right)$, where we have defined the momentum space projection operator as $P_g(\boldsymbol{k}) \equiv |g,\boldsymbol{k}\rangle\langle g,\boldsymbol{k}|$. Consequently, if initially the atoms have a continuous distribution in the momentum space, the final distribution of the ground state population will form fringes in the momentum space. Next, we will discuss what the fringes in the momentum space mean in the coordinate space.

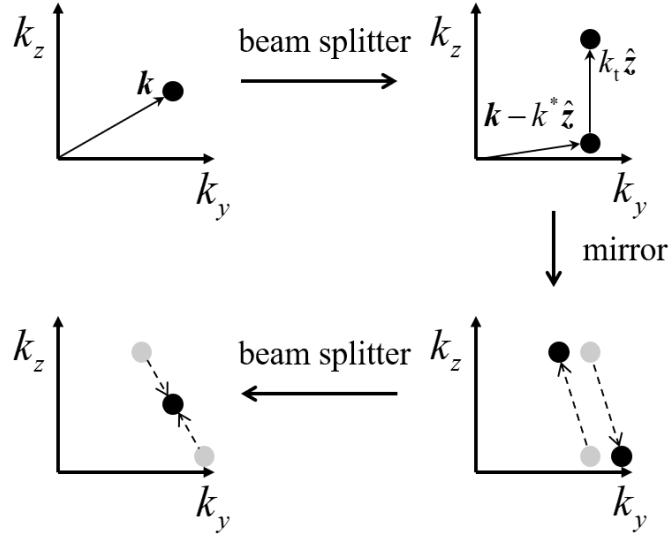

Fig. 2 The evolution of a momentum eigenstate $|\boldsymbol{k}\rangle$ in the $k-$space in a PSI under rotation. A $k-$eigenstate is a single dot in the $k-$space. The first atomic beam splitter splits $|\boldsymbol{k}\rangle$ into a superposition of two $k-$eigenstates separated by the momentum transfer $\boldsymbol{k}_t$. The atomic mirror switches the position of the two $k-$eigenstates in the absence of rotation. In the presence of rotation, however, the two $k-$eigenstates will be shifted in the $k_y-$direction. The second beam splitter will combine the two $k-$eigenstates and make them interfere.

Here we consider both the cases where the atoms are in a pure state and in a mixed state. If the atoms are in a pure state, then the external motion can be described by a wavefunction $\psi_k(\mathbf{k})$ for each atom. In the absence of rotation, the final external state of the atom internally in the ground state will be $\psi_k(\mathbf{k})\exp(-i\hbar \mathbf{k}^2(2T)/2m)$. According to the former discussion, in the presence of rotation, the final external state of the atom internally in the ground state will be

$$\psi'_k(\mathbf{k}) = \psi_k(\mathbf{k})\exp\left(-i\hbar \frac{\mathbf{k}^2}{2m}2T\right)\cos(\mathbf{r}_\Omega \cdot \mathbf{k}/2) \tag{1}$$

Rearranging Eq. (1) and eliminating the common phase factor, we can write:

$$\psi'_k(\mathbf{k}) = \frac{1}{2}\psi_k(\mathbf{k})\left[\exp\left(-i\hbar\frac{(\mathbf{k}+\mathbf{k}_1)^2}{2m}2T\right) + \exp\left(-i\hbar\frac{(\mathbf{k}-\mathbf{k}_1)^2}{2m}2T\right)\right] \tag{2}$$

where $\mathbf{k}_1 \equiv m\mathbf{r}_\Omega/4\hbar T = \mathbf{k}_\Omega/2$. To find the wavefunction of the atoms in the coordinate space, we compute the Fourier transform of $\psi'_k(\mathbf{k})$, to get:

$$\begin{aligned}\psi_r(\mathbf{r}) &= \mathcal{F}\left[\psi'_k(\mathbf{k})\right] \\ &= \frac{1}{2}\mathcal{F}\left[\psi_k(\mathbf{k})\exp\left(-i\hbar\frac{(\mathbf{k}+\mathbf{k}_1)^2}{2m}2T\right)\right] + \frac{1}{2}\mathcal{F}\left[\psi_k(\mathbf{k})\exp\left(-i\hbar\frac{(\mathbf{k}-\mathbf{k}_1)^2}{2m}2T\right)\right] \\ &= \frac{1}{2}\mathcal{F}\left[\psi_k(\mathbf{k}-\mathbf{k}_1)\exp\left(-i\hbar\frac{\mathbf{k}^2}{2m}2T\right)\right]e^{-i\mathbf{k}_1\cdot\mathbf{r}} + \frac{1}{2}\mathcal{F}\left[\psi_k(\mathbf{k}+\mathbf{k}_1)\exp\left(-i\hbar\frac{\mathbf{k}^2}{2m}2T\right)\right]e^{i\mathbf{k}_1\cdot\mathbf{r}} \\ &\equiv \frac{1}{2}\left[\psi_-(\mathbf{r})e^{-i\mathbf{k}_1\cdot\mathbf{r}} + \psi_+(\mathbf{r})e^{i\mathbf{k}_1\cdot\mathbf{r}}\right]\end{aligned} \tag{3}$$

where $\mathcal{F}$ stands for Fourier transform. The spatial distribution of the ground state $\langle P_g(\mathbf{r})\rangle = |\psi_r(\mathbf{r})|^2$ for an arbitrary $\psi_k(\mathbf{k})$ has to be calculated individually.

However, under the condition where the width of $\psi'_k(\mathbf{k})$ is much larger than $|\mathbf{k}_1|$ so that $\psi_-(\mathbf{r}) \approx \psi_+(\mathbf{r})$, both $\psi_-(\mathbf{r})$ and $\psi_+(\mathbf{r})$ will approximately equal $\mathcal{F}\left[\psi_k(\mathbf{k})\exp(-i\hbar \mathbf{k}^2 T/2m)\right]$, which is just the final external state of the atom internally in the ground state in the absence of rotation, as discussed before Eq. (1). In that case, $\langle P_g(\mathbf{r})\rangle$ is simply the product of the final profile of the atom cloud and a sinusoidal function $(1+\cos\mathbf{k}_\Omega\cdot\mathbf{r})/2$. This is exactly the result predicted by the conventional model.

The condition $\psi_-(\mathbf{r}) \approx \psi_+(\mathbf{r})$, corresponding to a smaller difference between $\psi_k(\mathbf{k}-\mathbf{k}_1)$ and $\psi_k(\mathbf{k}+\mathbf{k}_1)$, yields the highest contrast in the spatial interference fringes. A state wider in the momentum space corresponds to a smaller difference between $\psi_k(\mathbf{k}-\mathbf{k}_1)$ and $\psi_k(\mathbf{k}+\mathbf{k}_1)$. This condition also corresponds to a state narrower in the position space. Therefore, for a pure state, the narrower it is in the position space, the higher the contrast is for the spatial fringes. The limiting case of narrow wavefunctions in the position space is, of course, the point source.

However, the centers of mass of all trapped atoms cannot generally be described as a pure state. According to quantum statistical mechanics, the state of the center of mass of each atom can be described by a density operator $\rho = e^{-H/k_B T_K}$, where $H$ is the Hamiltonian, $k_B$ is the Boltzmann constant, and $T_K$ is the temperature. If we assume the atoms to be non-interacting and freely moving, we have $H = (\hbar k)^2 / 2m$, so that the state of the center of mass of each atom is described by a density operator $\rho = \int d\boldsymbol{k} \exp\left[-(\hbar \boldsymbol{k})^2 / 2m k_B T\right] |\boldsymbol{k}\rangle\langle \boldsymbol{k}|$. This density operator lacks coherence between different momentum eigenstates because these are also the eigenstates of energy. For such a system, there will be no spatial fringes at all. To see why, we recall that, for a pure state, the width in the $k$–space determines the contrast of the spatial interference fringes. Every pure state in the density operator $\rho$ has no width in the $k$–space. Consequently, no spatial fringe will appear. The existence of coherence between different $|k\rangle$ states for atoms cooled by lasers have been demonstrated in experiments[12,13,14]. Therefore, the diagonal density matrix is inadequate, and we need a different model to describe the initial state of such cold atoms.

We consider a situation where the atoms released from a magneto-optic trap is caught in an isotropic dipole force trap before the onset of the PSI process. Such a trap can be modeled as a harmonic potential well[15] with a characteristic frequency $\omega$, so that the Hamiltonian can be expressed as:

$$H = \frac{(\hbar \boldsymbol{k})^2}{2m} + \frac{1}{2} m \omega^2 \boldsymbol{x}^2 \tag{4}$$

The energy eigenstates of this Hamiltonian are

$$|n\rangle = \frac{a}{\sqrt{\pi^{1/2} 2^n n!}} \int d\boldsymbol{k} H_n(ka) e^{-(ka)^2/2} |\boldsymbol{k}\rangle \tag{5}$$

where $a = \sqrt{\hbar/m\omega}$ is a measure of the size of the trap and $H_n$ is the $n$ th order Hermite polynomial. The density operator of the atoms in this case can be expressed as:

$$\rho = \sum_{n=0}^{\infty} \exp\left[-\frac{[\hbar\omega(n+1/2)]^2}{k_B T_K}\right] |n\rangle\langle n| \tag{6}$$

During the expansion of the atom cloud, upon release from the trap, each $|n, g\rangle$ (defined as the state where the external state is $|n\rangle$ and the internal state is $|g\rangle$) evolves independently. The evolution of each $|n, g\rangle$ under the sequence of pulses used for the PSI can also be evaluated individually. We define as $\psi_n(\mathbf{k})$ the final external state corresponding to $|n,g\rangle$, for an atom in the ground state internally. The signal at the end of the PSI process can thus be expressed as:

$$\langle P_g \rangle = \text{tr}(\rho P_g) = \sum_{n=0}^{\infty} \exp\left[-\frac{[\hbar\omega(n+1/2)]^2}{k_B T_K}\right] |\psi_n(\mathbf{k})^2| \qquad (7)$$

where the projection operator is defined as $P_g \equiv |g\rangle\langle g|$. Eq. (8) shows that the overall contrast is determined by the sum of the fringes resulting from each energy eigenstate, $|n,g\rangle$. Therefore, the smaller $a$ is, the narrower all the energy eigenstates will be in position space, and the higher the contrast of the spatial fringes will be.

### 4. Large momentum transfer by additional Raman pulses

LMT atom optics are broadly defined as methods that increase the momentum splitting between the interferometer arms beyond 2ℏk. In light-pulse atom interferometry, several LMT techniques have been demonstrated. These include using an additional sequence of π pulses[5,6,16,17,18,19,20] or Bloch oscillations in an optical lattice[21,22,23,24] following the initial π/2 pulse to increase the momentum splitting, as well as implementing individual π/2 pulses that transfer an increased number of photon momentum recoils via higher order Bragg diffraction[25]. For the sequential pulse method, either Raman transitions[16], which change the internal hyperfine state, or Bragg transitions[5,6,19,20], which leave the internal state unchanged, can be used. Both methods have their advantages and are worth considering for a given application. For instance, Raman transitions are capable of efficiently transferring atom clouds with wider velocity spreads along the laser beam axis[26], while Bragg transitions are immune to sources of noise or drift arising from effects involving a changing internal state, such as ac Stark shifts of the transition resonance[8,17,25,27]. Bloch oscillations also have the advantage of very high momentum transfer efficiency that is robust against intensity inhomogeneities across the atom cloud[21,22,23,24]. Sequences of single-photon transitions on the 689 nm inter-combination transition of strontium[28] represent an alternative approach that offers wide velocity acceptance and reduced AC Stark shifts. This promising approach will be studied in future work.

Techniques such as Bragg diffraction and Bloch oscillation in optical lattices require atoms with sub-recoil velocity spreads in the longitudinal direction, requiring either velocity selection or increased cooling, which adversely affect the signal to noise ratio as well as the repetition rate. As such, we focus here on the method of using additional Raman pulses[16,17,18]. The protocol for realizing LMT using this method is illustrated in Fig. 3. Additional Raman pulses in alternating directions are added to the conventional $\pi/2 - \pi - \pi/2$ pulse sequence.

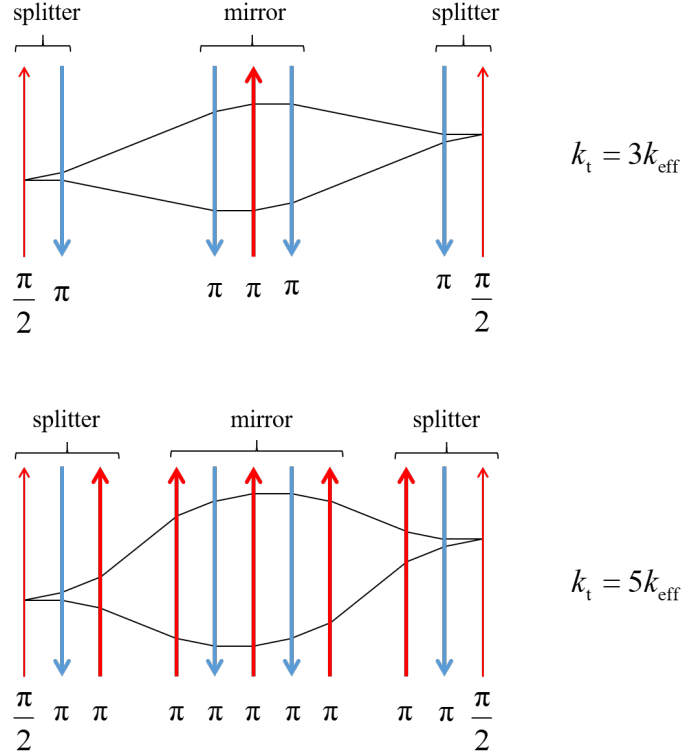

Fig. 3 Large momentum transfer by additional $\pi-$ pulses in alternating directions. Each Raman pulse here is a pair of counterpropagating Raman beams. The arrows mean the effective directions of Raman pulses.

The modeling of the motion of the center of mass of each atom was discussed earlier in Section 3. Here, we describe the evolution of the internal states of each atom under this Raman pulse sequence. The internal state is modeled as a three level system: the ground state $|g\rangle$, the excited state $|e\rangle$, and the intermediate state $|i\rangle$. The pulses induce Raman transitions among these three states. The frequency and the wavenumber of the first (second) Raman beam are denoted as $\omega_1$ ($\omega_2$) and $\boldsymbol{k}_1$ ($\boldsymbol{k}_2$). Due to conservation of linear momentum, a pair of Raman beams couples the three states $|g,\boldsymbol{k}\rangle$, $|i,\boldsymbol{k}+\boldsymbol{k}_1\rangle$, and $|e,\boldsymbol{k}+\boldsymbol{k}_1-\boldsymbol{k}_2\rangle$. The resulting Hamiltonian, in the basis spanned by these three states, can be expressed as follows:

$$H_{\text{Raman}} = \hbar \begin{bmatrix} \dfrac{\hbar \boldsymbol{k}^2}{2m} + \dfrac{\delta_0}{2} & \dfrac{\Omega_1}{2} & 0 \\ \dfrac{\Omega_1}{2} & \dfrac{\hbar(\boldsymbol{k}+\boldsymbol{k}_1)^2}{2m} - \Delta_0 & \dfrac{\Omega_2}{2} \\ 0 & \dfrac{\Omega_2}{2} & \dfrac{\hbar(\boldsymbol{k}+\boldsymbol{k}_1-\boldsymbol{k}_2)^2}{2m} - \dfrac{\delta_0}{2} \end{bmatrix} \quad (8)$$

where $\delta_0 = \delta_{g0} - \delta_{e0}$ and $\Delta_0 = (\delta_{g0} + \delta_{e0})/2$. Here, $\delta_{g0}$ is defined as $\omega_1 - (\omega_i - \omega_g)$ and $\delta_{e0}$ as $\omega_2 - (\omega_i - \omega_e)$. Next, we carry out the adiabatic elimination[29], corresponding to making the approximation that

$$i\dot{c}_i = \frac{\Omega_1}{2}c_g + \left[\frac{\hbar(k+k_1)^2}{2m} - \Delta_0\right]c_i + \frac{\Omega_2}{2}c_e \approx 0 \tag{9}$$

Substituting $c_i$ solved from Equation (10) into the Schrödinger equation, we have

$$i\dot{c}_g = \left\{\frac{\hbar k^2}{2m} + \frac{\delta_0}{2} - \frac{\Omega_1^2}{2\left[\frac{\hbar(k+k_1)^2}{m} - 2\Delta_0\right]}\right\}c_g - \frac{\Omega_1\Omega_2}{2\left[\frac{\hbar(p+k_1)^2}{m} - 2\Delta_0\right]}c_e$$

$$i\dot{c}_e = \left\{\frac{\hbar(k+k_1-k_2)^2}{2m} - \frac{\delta_0}{2} - \frac{\Omega_2^2}{2\left[\frac{\hbar(k+k_1)^2}{m} - 2\Delta_0\right]}\right\}c_e - \frac{\Omega_1\Omega_2}{2\left[\frac{\hbar(k+k_1)^2}{m} - 2\Delta_0\right]}c_g \tag{10}$$

Therefore, the effective two-level system Hamiltonian, in the basis spanned by states $|g,k\rangle$ and $|e,k+k_1-k_2\rangle$, is

$$H = \begin{bmatrix} \frac{\hbar k^2}{2m} + \frac{\delta_0}{2} - \frac{\Omega_1^2}{2\left[\frac{\hbar(k+k_1)^2}{m} - 2\Delta_0\right]} & \frac{-\Omega_1\Omega_2}{2\left[\frac{\hbar(k+k_1)^2}{m} - 2\Delta_0\right]} \\ \frac{-\Omega_1\Omega_2}{2\left[\frac{\hbar(k+k_1)^2}{m} - 2\Delta_0\right]} & \frac{\hbar(k+k_1-k_2)^2}{2m} - \frac{\delta_0}{2} - \frac{\Omega_2^2}{2\left[\frac{\hbar(k+k_1)^2}{m} - 2\Delta_0\right]} \end{bmatrix} \tag{11}$$

If we set $\delta_0 = \hbar(k_1-k_2)^2/2m$, $\Omega_1 = \Omega_2 \equiv \Omega_0$, and shift all energy levels by an amount that makes the energy of state $|g,k\rangle$ vanish, we have

$$H = \hbar \begin{bmatrix} 0 & \dfrac{\Omega_0^2}{4\tilde{\Delta}_0} \\ \dfrac{\Omega_0^2}{4\tilde{\Delta}_0} & \dfrac{\hbar \boldsymbol{k} \cdot (\boldsymbol{k}_1 - \boldsymbol{k}_2)}{m} \end{bmatrix} \qquad (12)$$

where $\tilde{\Delta}_0 \equiv \left[\Delta_0 - \hbar(\boldsymbol{k}+\boldsymbol{k}_1)^2/2m\right]$. From this Hamiltonian we see a detuning caused by the Doppler shift given by $\hbar \boldsymbol{k} \cdot (\boldsymbol{k}_1 - \boldsymbol{k}_2)/m$, as well as the effective Rabi frequency given by $\Omega_0^2/2\tilde{\Delta}_0$. In addition, the adiabatic elimination also gives us the effective decay rate[29] between the states $|g\rangle$ and $|e\rangle$. Using Eq. (10), for $\Omega_1 = \Omega_2 \equiv \Omega_0$, the population of the intermediate state $|i\rangle$ is given by:

$$|c_i|^2 \approx \frac{\Omega_0^2}{4\tilde{\Delta}_0^2}\left[\{c_g^*\langle g| + c_e^*\langle e|\}\{c_g|g\rangle + c_e|e\rangle\}\right] = \frac{\Omega_0^2}{4\tilde{\Delta}_0^2}\left(|c_g|^2 + |c_e|^2\right) \approx \frac{\Omega_0^2}{4\tilde{\Delta}_0^2} \qquad (13)$$

Therefore, the effective decay rates are

$$\Gamma_{g\to e}^{\text{eff}} = \frac{\Gamma_{i\to e}\Omega_0^2}{4\tilde{\Delta}_0^2}; \qquad \Gamma_{e\to g}^{\text{eff}} = \frac{\Gamma_{i\to g}\Omega_0^2}{4\tilde{\Delta}_0^2} \qquad (14)$$

The total decay rate for the coherence between states $|g\rangle$ and $|e\rangle$ is then given by:

$$\Gamma_{\text{eff}} = \Gamma_{g\to e}^{\text{eff}} + \Gamma_{e\to g}^{\text{eff}} = \frac{\Gamma\Omega_0^2}{4\tilde{\Delta}_0^2} \qquad (15)$$

The D$_2$ line decay rate from $5^2P_{3/2}$ to $5^2S_{1/2}$, expressed as $\Gamma = \Gamma_{i\to g} + \Gamma_{i\to e}$, is about $2\pi \times (6\text{ MHz})$ for $^{87}$Rb. Only the atoms that have not experienced spontaneous emission have kept their phase information. The fraction of atoms that have decohered by the end of the interferometry process is given by $\exp(-\Gamma_{\text{eff}}\tau)$, where $\tau$ is the total duration of all the Raman pulses. With this model, we can simulate the signal for a PSI-LMT, while taking into account the complexities caused by detuning. The effect of spontaneous emission is taken into account later on in a heuristic manner. In the model discussed in §3, we assumed all $k-$ components to be resonant, which is approximately valid if the effective Rabi frequency is much larger than the Doppler shift. In order to account for more general conditions, in our simulation we use different Hamiltonian operators for different $k-$ components, corresponding to Eq. (13). To determine quantitatively the density of fringes, we compute the Fourier transform of the pattern. Experimentally, this Fourier transform can be done in real time using a lens in the system for imaging the atom cloud. Thus, our signal is expressed as $\tilde{P}_g(\tilde{\boldsymbol{k}}) = \int d\boldsymbol{r}\, e^{-i\tilde{\boldsymbol{k}}\cdot\boldsymbol{r}} \langle P_g(\boldsymbol{r})\rangle$, where $P_g(\boldsymbol{r})$ is the position space projection operator for atoms in the ground state, as defined earlier. It should be noted that $\tilde{P}_g(\tilde{\boldsymbol{k}})$ is different from $P_g(\boldsymbol{k})$, the momentum space projection operator for atoms in the ground state, also defined earlier.

For a pure state under the condition that $\psi_-(r) \approx \psi_+(r) \equiv \psi_0(r)$ and in the limit that $\Omega_0^2/2\Delta_0 \to \infty$, the signal can be expressed as:

$$\tilde{P}_g(\tilde{k}) = \mathcal{F}\left[\frac{1}{2}|\psi_0(r)|^2 (1+\cos k_\Omega \cdot r)\right] = \frac{1}{2}f(\tilde{k}) + \frac{1}{4}f(\tilde{k}-k_\Omega) + \frac{1}{4}f(\tilde{k}+k_\Omega) \quad (16)$$

where $f(\tilde{k}) \equiv \mathcal{F}\left[|\psi_0(r)|^2\right]$. The spatial fringes representing $\langle P_g(r) \rangle$ and the corresponding Fourier transforms given by $\tilde{P}_g(\tilde{k})$ for such an ideal case is depicted in Fig. 4. The left panel shows plots for $k_t = k_{\text{eff}}$ and the right panel show plots for $k_t = 3k_{\text{eff}}$. In each panel, (a) is the plot of $\langle P_g(r) \rangle$ in the plane perpendicular to $k_t$, (b) is the cross section at the dashed line in (a), (c) is the plot of $\tilde{P}_g(\tilde{k})$ in the plane perpendicular to $k_t$, and (d) is the cross section at the dashed line of (c).

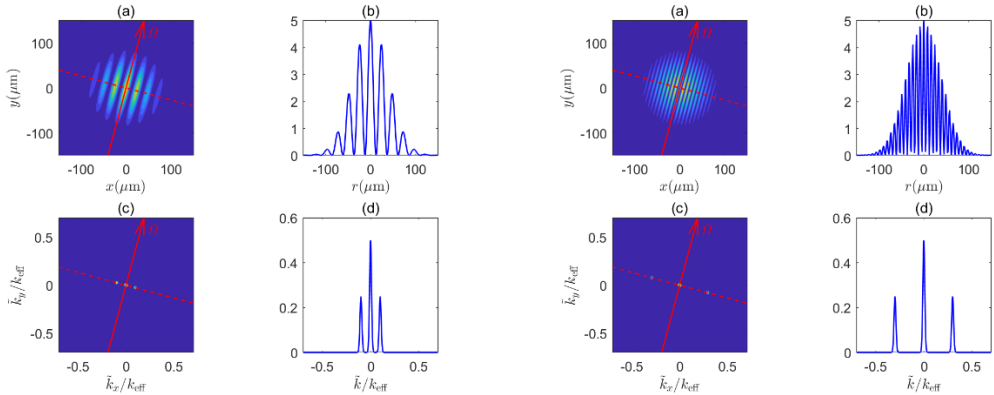

Fig. 4 Signals for the conventional PSI and the PSI-LMT without considering the detuning effect. The left panel corresponds to $k_t = k_{\text{eff}}$ and the right panel corresponds to $k_t = 3k_{\text{eff}}$. In each panel, (a) is the plot of $\langle P_g(r) \rangle$ in the plane perpendicular to $k_t$, (b) is the cross section at the dashed line in (a), (c) is the plot of $\tilde{P}_g(\tilde{k})$ in the plane perpendicular to $k_t$, and (d) is the cross section at the dashed line of (c). The orientation of the signal indicates the direction of the angular velocity.

We have simulated the signals for the case of a harmonic oscillator trap, as shown in Eq. (7), for $^{87}\text{Rb}$, with the following parameters: $\Omega_0 = 2\pi \times (10\sqrt{10} \text{ MHz})$, $\Delta_0 = 2\pi \times (500 \text{ MHz})$, $T_K = 6 \text{ μK}$, and $a = 0.1 \text{ μm}$. Here, we have chosen an unrealistically small size of the trap, in order to elucidate the behavior of a system that is very close to an ideal point source. The simulation results for this case is shown in Fig. 5. The main difference from the result shown in Fig. 4 is that the height of the signal peak is shorter, due to the fact that the detuning resulting from $k_t$ is taken into account. There is also a little difference in the width of the signal peak. For the LMT case shown in the right panel, there are also some small peaks in addition to the main signal peak. This is because the pulses are not ideal. For example, a pulse that is nominally designate to be a $\pi-$pulse, does not fully transform a ground state to an excited state, or vice versa, but will leave some residual. The small peaks are the consequence of the interference involving the residuals.

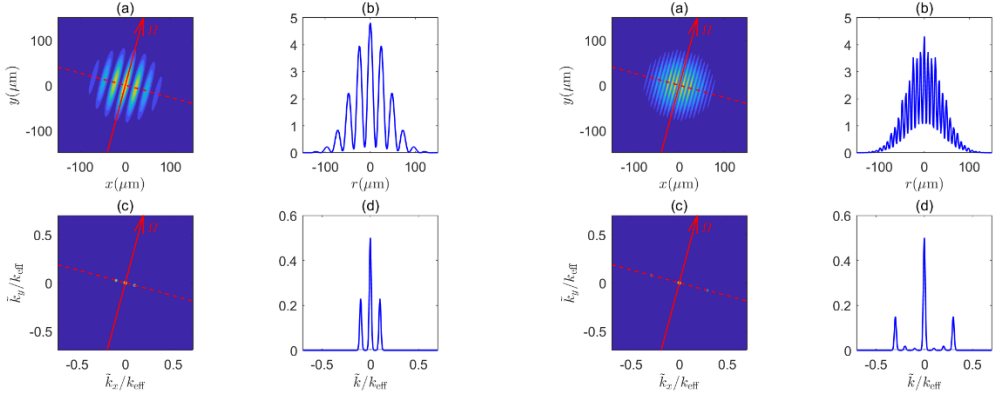

Fig. 5 Simulation results for the conventional PSI (left panel: $k_\mathrm{t} = k_\mathrm{eff}$) and the LMT-PSI (right panel: $k_\mathrm{t} = 3k_\mathrm{eff}$) employing $^{87}\mathrm{Rb}$ with the parameters $\Omega_0 = 2\pi \times \left(10\sqrt{10}\ \mathrm{MHz}\right)$, $\Delta_0 = 2\pi \times \left(500\ \mathrm{MHz}\right)$, $T_\mathrm{K} = 6\ \mu\mathrm{K}$, and $a = 0.1\ \mu\mathrm{m}$, including the detuning effect. In each panel, (a) is the plot of $\langle P_g(\mathbf{r}) \rangle$ in the plane perpendicular to $\mathbf{k}_\mathrm{t}$, (b) is the cross section at the dashed line in (a). (c) is the plot of $\tilde{P}_g(\tilde{\mathbf{k}})$ in the plane perpendicular to $\mathbf{k}_\mathrm{t}$, and (d) is the cross section at the dashed line of (c). The orientation of the signal indicates the direction of the angular velocity.

As can be seen from comparisons between the PSI and LMT-PSI results shown in Fig. 5, the PMT process produces a larger separation for the signal peaks in the Fourier transform domain, thus making it more sensitive for measuring rotation. At the same time, the amplitudes of the signal peaks are smaller, which in turn would represent a reduction in the effective signal to noise ratio, and a corresponding reduction in sensitivity. The actual improvement in sensitivity would be determined by both factors. In this context, we consider first the fact that the degradation of the signal (both in terms of the reduction in the amplitudes, and the appearance of additional peaks) can be countered by increasing the effective Rabi frequency: $\Omega_0^2/2\Delta_0$. As we mentioned before, the maximum heights for the signal peaks occur when $\Omega_0^2/2\Delta_0 \to \infty$. Fig. 6 shows the comparison between signals for different effective Rabi frequencies, with $k_\mathrm{t} = 3k_\mathrm{eff}$. The left panel in Fig. 6 corresponds to the case where $\Omega_0 = 2\pi \times \left(10\sqrt{10}\ \mathrm{MHz}\right)$ and $\Delta_0 = 2\pi \times \left(500\ \mathrm{MHz}\right)$. The right panel corresponds to the case where $\Omega_0 = 2\pi \times \left(100\ \mathrm{MHz}\right)$ and $\Delta_0 = 2\pi \times \left(500\ \mathrm{MHz}\right)$. As can be seen, the amplitudes of the signal peaks increase for the larger value of the effective Rabi frequency, and the additional peaks almost disappear completely. Fig. 7 shows a signal for $k_\mathrm{t} = 7k_\mathrm{eff}$ with $\Omega_0 = 2\pi \times \left(100\ \mathrm{MHz}\right)$ and $\Delta_0 = 2\pi \times \left(500\ \mathrm{MHz}\right)$. We can see in this case the signal is still very close to the ideal signal that does not take into account the effect of detuning. Simulations with larger $k_t$ requires too much computational resources because we are using a fully quantum model. Attempts will be made in the near future to extend the simulation to much larger values of $k_t$. In what follows, we present a systematic analysis for estimating quantitatively the expected net enhancement in sensitivity, as a function of the effective Rabi frequency and the value of $k_t$, while taking into account the effect of spontaneous emission heuristically.

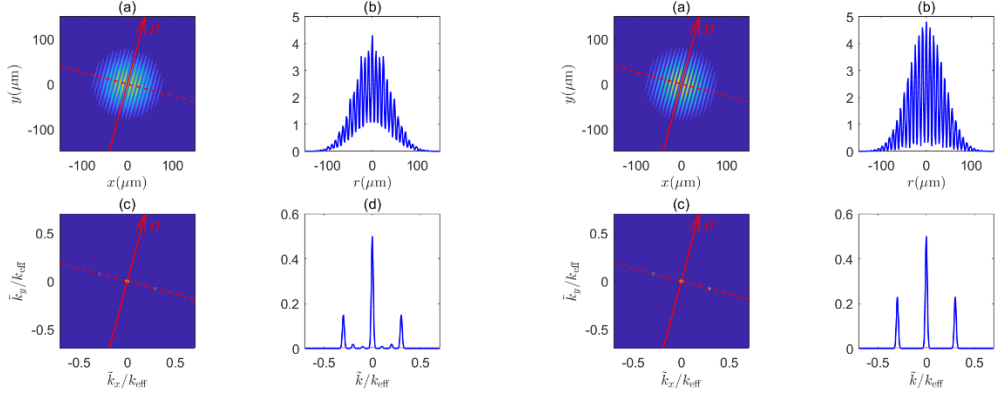

Fig. 6 Comparison between the signals with low and high effective Rabi frequencies, with $k_t = 3k_{eff}$. The left panel corresponds to the case where $\Omega_0 = 2\pi \times (10\sqrt{10}\text{ MHz})$ and $\Delta_0 = 2\pi \times (500\text{ MHz})$. The right panel corresponds to the case where $\Omega_0 = 2\pi \times (100\text{ MHz})$ and $\Delta_0 = 2\pi \times (500\text{ MHz})$. With the higher effective Rabi frequency, the contrast of the signal is improved significantly.

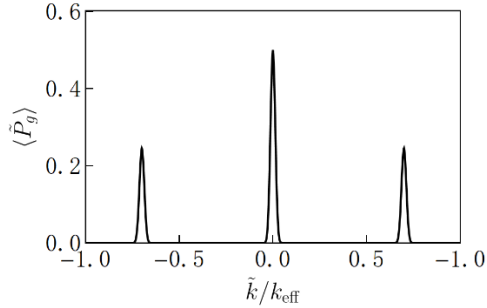

Fig. 7 The signal for $k_t = 7k_{eff}$ with $\Omega_0 = 2\pi \times (100\text{ MHz})$ and $\Delta_0 = 2\pi \times (500\text{ MHz})$. In this case, the signal is still very close to the ideal signal that does not take into account the effect of detuning.

Assuming that $k_t$ is perpendicular to $\Omega$, we have $k_\Omega = k_t \Omega T$. Therefore, the uncertainty of $k_\Omega$ determines the uncertainty of $\Omega$ according to the relation $\delta\Omega = \delta k_\Omega / k_t T$. We denote by $h$ the amplitude of the signal peak in the Fourier transform domain. In general, the uncertainty of a signal is the linewidth divided by the signal-to-noise ratio. Using this rule, we can write that $\delta k_\Omega = \alpha \gamma / \sqrt{h}$, where $\gamma$ is the width of the signal peak, and $\alpha$ is a constant coefficient. It then follows that $(\delta\Omega)^{-1} = k_t T \sqrt{h}/\alpha\gamma$. Since the signal is in the Fourier transform domain, $\gamma$ is approximately the inverse of the final size of the atomic cloud. Here, we do not consider the inhomogeneity of the Raman beams. As a result, $\gamma$ will only be determined by the free expansion but not affected significantly by the LMT process. The effect of a more realistic Raman beam intensity profile will be investigated in the future. Thus, we see that the larger the final atomic cloud size is, the smaller $\gamma$ is, and the smaller $\delta\Omega$ is. To compare LMT-PSI's with different values of $k_t = Nk_{\text{eff}}$, we make their atomic clouds end up with the same final size, and therefore the same $\gamma$. We define an improvement parameter $\varepsilon \equiv \delta\Omega_{Nk_{\text{eff}}}/\delta\Omega_{k_{\text{eff}}}^{\text{ideal}} = N\sqrt{h_{Nk_{\text{eff}}}/h^{\text{ideal}}}$, where the ideal case is produced when the effective Rabi frequency for stationary atoms $\Omega_0^2/2\Delta_0 \to \infty$ and there is no effective spontaneous emission, as noted earlier.

The value of $h$ is determined primarily by the transition efficiency of each $\pi-$pulse and the effective spontaneous emission. To simplify the analysis, we ignore the dependence of the effective Rabi frequency on the momentum of the atoms. Therefore, we define the constant effective Rabi frequency as $\Omega_{\text{eff}} \equiv \Omega_0^2/2\Delta_0$. We define the propagator of the quantum state of the atom due to a Raman pulse, $U$, by the expression $|\psi(t_0+t)\rangle = U(t)|\psi(t_0)\rangle$. This propagator can be expressed as:

$$U(t) = \begin{bmatrix} \cos\frac{\Omega' t}{2} - i\frac{\delta}{\Omega'}\sin\frac{\Omega' t}{2} & -i\frac{\Omega_{\text{eff}}}{\Omega'}\sin\frac{\Omega' t}{2} \\ -i\frac{\Omega_{\text{eff}}}{\Omega'}\sin\frac{\Omega' t}{2} & \cos\frac{\Omega' t}{2} - i\frac{\delta}{\Omega'}\sin\frac{\Omega' t}{2} \end{bmatrix} \quad (17)$$

where $\Omega' \equiv \sqrt{\Omega_{\text{eff}}^2 + \delta^2}$ and $\delta$ is the detuning caused by the Doppler shift. In principle, even the atoms following the same trajectories will have a thermal distribution of momenta. However, in the LMT case with $N$ much larger than unity, the thermal momentum is very small compared to $k_t$. With $T_K = 6$ μK, the typical thermal momentum, $\sqrt{mk_B T_K}$, is only $\sim 2\hbar k_{\text{eff}}$. Therefore, we ignore the thermal momentum distribution of the atoms, which means that atoms following the same trajectories experience the same detuning. Generally, it is difficult to handle this propagator analytically. However, in the limit that $\delta \ll \Omega_{\text{eff}}$, we can make approximations to Eq. (18) and make it more manageable. The transition efficiency of a $\pi$ pulse derived from Eq. (18) is

$$\eta_k = \left(\frac{\Omega_{\text{eff}}}{\Omega'}\sin\frac{\Omega' t}{2}\right)^2 = \frac{1}{1+\delta_k^2/\Omega_0^2}\sin^2\frac{\mu\sqrt{1+\delta_k^2/\Omega_{\text{eff}}^2}}{2}$$

$$\equiv \frac{1}{1+\beta_k}\sin^2\frac{\mu\sqrt{1+\beta_k}}{2} = \frac{1}{1+\beta_k} \quad (18)$$

where $\mu \equiv \Omega_{\text{eff}} t$, $\delta_k$ is the Doppler shift inducing detuning for an atom with momentum $k$, and $\beta_k \equiv (\delta_k/\Omega_{\text{eff}})^2 = (\hbar k_{\text{eff}} k/m\Omega_{\text{eff}})^2$. In the last step of Eq. (19), we have assumed that we can make $\mu$ different for each Raman pulse, such that for all pulses $\mu_k = \frac{\pi}{\sqrt{1+\beta_k}}$. When $\Omega_{\text{eff}} = 2\pi \times (10 \text{ MHz})$, we have $\beta_{100 k_{\text{eff}}}$ is approximately only 0.1. Therefore, it is reasonable to consider $\beta_k$ a small quantity. The height of the signal peak for $k_t = N k_{\text{eff}}$ is

$$h = \frac{1}{4} \left( \eta_{k_{\text{eff}}} \eta_{2k_{\text{eff}}} \cdots \eta_{(N-1)k_{\text{eff}}/2} \right)^4 \exp(-\Gamma_{\text{eff}} \tau)$$

$$= \frac{1}{4\left[ \left(1+\beta_{k_{\text{eff}}}\right)\left(1+\beta_{2k_{\text{eff}}}\right)\cdots\left(1+\beta_{(N-1)k_{\text{eff}}/2}\right) \right]^4} \quad (19)$$

$$\exp\left( -\Gamma_{\text{eff}} \frac{\pi}{\Omega_{\text{eff}}} 4 \left( \frac{1}{\sqrt{1+\beta_{k_{\text{eff}}}}} + \frac{1}{\sqrt{1+\beta_{2k_{\text{eff}}}}} + \cdots \frac{1}{\sqrt{1+\beta_{(N-1)k_{\text{eff}}/2}}} \right) \right)$$

Then the logarithm of the height is:

$$\ln h = -4 \sum_{n=1}^{(N-1)/2} \left( \ln(1+\beta_{nk_{\text{eff}}}) + \frac{\pi \Gamma_{\text{eff}}}{\Omega_{\text{eff}} \sqrt{1+\beta_{nk_{\text{eff}}}}} \right) - 2\ln 2 \quad (20)$$

Keeping only the leading term of $\beta$, we have:

$$\ln h \approx -4 \left( 1 - \frac{\pi \Gamma_{\text{eff}}}{2\Omega_{\text{eff}}} \right) \left( \frac{\hbar k_{\text{eff}}^2}{m\Omega_{\text{eff}}} \right)^2 \sum_{n=1}^{(N-1)/2} n^2 - \frac{2(N-1)\pi \Gamma_{\text{eff}}}{\Omega_{\text{eff}}} - 2\ln 2$$

$$= -\frac{1}{6} N(N^2-1) \left( 1 - \frac{\pi \Gamma_{\text{eff}}}{2\Omega_{\text{eff}}} \right) \left( \frac{\hbar k_{\text{eff}}^2}{m\Omega_{\text{eff}}} \right)^2 - \frac{2(N-1)\pi \Gamma_{\text{eff}}}{\Omega_{\text{eff}}} - 2\ln 2 \quad (21)$$

$$\approx -\frac{1}{6} N^3 \left( 1 - \frac{\pi \Gamma_{\text{eff}}}{2\Omega_{\text{eff}}} \right) \left( \frac{\hbar k_{\text{eff}}^2}{m\Omega_{\text{eff}}} \right)^2 - \frac{2\pi N \Gamma_{\text{eff}}}{\Omega_{\text{eff}}} - 2\ln 2$$

Normally, as long as we have a reasonably large $\Delta_0$, the value of $\pi \Gamma_{\text{eff}}/\Omega_{\text{eff}}$ is very small compared to 1. For example, if $\Omega_0 = 2\pi \times (100 \text{ MHz})$ and $\Delta_0 = 2\pi \times (500 \text{ MHz})$, we have $\pi \Gamma_{\text{eff}}/\Omega_{\text{eff}} = 0.006$. Therefore, $(1 - \pi \Gamma_{\text{eff}}/2\Omega_{\text{eff}}) \approx 1$. With a reasonably large $\Delta_0$, we also note that $\Gamma_{\text{eff}} = \Gamma \Omega_0^2/4\tilde{\Delta}_0^2 \approx \Gamma \Omega_0^2/4\Delta_0^2 = \Gamma \Omega_{\text{eff}}/2\Delta_0$. Then we can calculate the natural logarithm of the improvement factor:

$$\ln \varepsilon = \ln N + \frac{1}{2} \ln \frac{h_{Nk_{\text{eff}}}}{h^{\text{ideal}}} = \ln N - \frac{1}{3} N^3 \left( \frac{\hbar k_{\text{eff}}^2 \Delta_0}{m\Omega_0^2} \right)^2 - \frac{\pi N \Gamma}{2\Delta_0} \quad (22)$$

The optimal value of $\Delta_0$ for maximizing $\ln \varepsilon$ is:

$$\Delta_0^{opt} = \left(\frac{3\pi\Gamma}{4}\right)^{1/3} \left(\frac{m\Omega_0^2}{N\hbar k_{eff}^2}\right)^{2/3} \qquad (23)$$

The natural logarithm of the improvement factor for this $\Delta_0$ is

$$\ln\varepsilon = \ln N - N^{5/3}\left(\frac{3\pi\Gamma\hbar k_{eff}^2}{4m\Omega_0^2}\right)^{2/3} \qquad (24)$$

We see that the maximum value of $\varepsilon$ is given by:

$$\varepsilon_{max} = e^{-3/5}\left(\frac{3}{125}\right)^{1/5}\left(\frac{4m\Omega_0^2}{\pi\Gamma\hbar k_{eff}^2}\right)^{2/5} = 0.56\left[\frac{\Omega_0}{2\pi\times(1\text{ MHz})}\right]^{4/5} \qquad (25)$$

This value of $\varepsilon$ occurs for an optimal value of $N$, given by:

$$N_{opt} = \left(\frac{3}{125}\right)^{1/5}\left(\frac{4m\Omega_0^2}{\pi\Gamma\hbar k_{eff}^2}\right)^{2/5} = 1.0\left[\frac{\Omega_0}{2\pi\times(1\text{ MHz})}\right]^{4/5} \qquad (26)$$

We can see from Equation (26) and (27) that both $\varepsilon_{max}$ and $N_{opt}$ are proportional to $\Omega_0^{4/5}$. Fig. 8 shows how $\varepsilon$ varies with $N$ for $\Omega_0 = 2\pi\times(200\text{ MHz})$ (red) and $\Omega_0 = 2\pi\times(100\text{ MHz})$ (blue). We see that $\varepsilon_{max} = 39$ for $N_{opt} = 69$ with $\Omega_0 = 2\pi\times(200\text{ MHz})$. With this Rabi frequency, when $N = 69$, the optimal $\Delta_0$ is $2\pi\times(1.7\text{ GHz})$, according to Eq. (24). It is also shown in Fig. 8 that $\Omega_0$ can significantly affect the maximum improvement LMT can achieve.

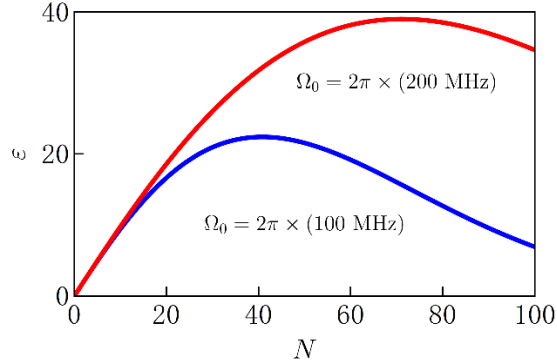

Fig. 8 Improvement factor $\varepsilon$ as a function of $N = k_t/k_{eff}$, with the effective Rabi frequency $\Omega_0 = 2\pi\times(200\text{ MHz})$ (red) and $\Omega_0 = 2\pi\times(100\text{ MHz})$ (blue). $\varepsilon$ reaches a maximum value of 39 for $N=69$ with $\Omega_0 = 2\pi\times(200\text{ MHz})$.

It can be seen from the discussion above that the value of $\Omega_0$ is very important for the performance of the LMT-PSI. Therefore, we discuss here the relation between the experimental parameters and $\Omega_0$. For $^{87}$Rb, we assume the ground state $|g\rangle$ to be $\{^2S_{1/2},\ F=1,\ m_F=0\}$, and the excited state $|e\rangle$ to be $\{^2S_{1/2},\ F=2,\ m_F=0\}$. In most implementations of Raman-pulse-based atom interferometers, the beams are circularly ($\sigma$) polarized[30]. If the beams are $\sigma^+$ polarized, the intermediate state $|i\rangle$ consists of two states: $|F=1,m_F=1\rangle$ and $|F=2,m_F=1\rangle$ of the $^2P_{3/2}$ manifold. The corresponding transition matrix elements[31] are shown in Figure 9. For the cycling transition from $|F=2,m_F=2\rangle$ to $|F=3,m_F=3\rangle$, an intensity of 3.34 mW/cm² yields $\Omega_0=\Gamma$. For a given intensity on each leg of the Raman transition, we can use this information to determine the effective Rabi frequency for each of the two Raman transitions, treated separately, and the net effective Rabi frequency would be the sum of these two effective Rabi frequencies. If we assume that each leg has the same laser intensity, and consider the fact that the energy separation between the two upper levels (~157 MHz) is negligible compared to the detuning, then it is easy to see that the effective Rabi frequency for the lower Raman transition is weaker than that for the upper Raman transition by a factor of $\left(\sqrt{1/8}\times\sqrt{1/8}\right)/\left(\sqrt{5/24}\times\sqrt{1/120}\right)=3$. If we consider the upper Raman transition only, the intensity needed for the condition of $\Omega_0=2\pi\times(100\text{ MHz})\approx 16.7\Gamma$ is ~3.7 W/cm². When both Raman transitions are taken into account, an intensity lower by a factor of 3/4 (i.e., ~2.8 W/cm²) would produce the effective Rabi frequency corresponding to $\Omega_0=16.7\Gamma$ in our model presented above[32]. Such an intensity can be achieved, for example, by using a tapered amplifier on each leg of the Raman transition.

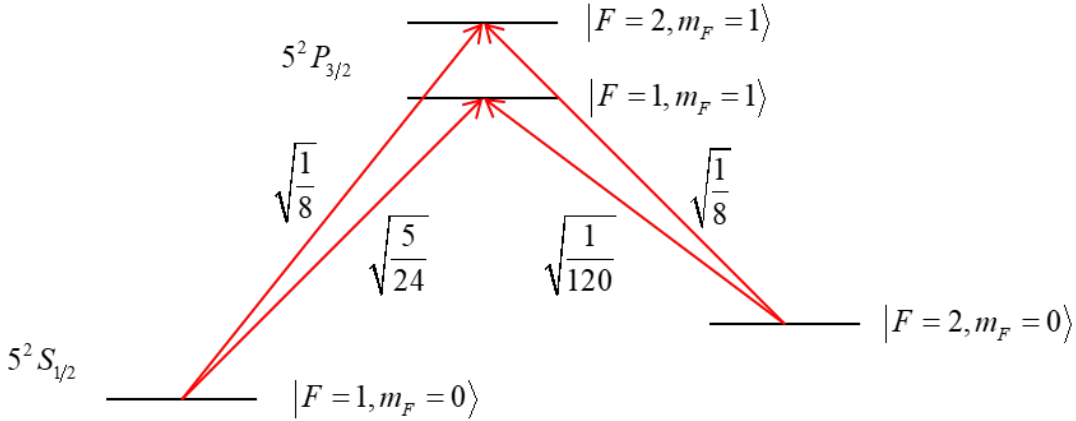

Fig. 9 Matrix elements relevant to the Raman transition between $|g\rangle$ and $|e\rangle$.

There is another technique that can potentially decrease the effect of the detuning. At the beginning, when the momentum difference between the two arms are small, both arms are addressed with the same Raman beams. When the momentum difference between the two arms become large enough, we can address them with different Raman beams so that both arms are resonant to its own Raman beams and far detuned from the Raman beams for the other arm. This technique works well for very cold atoms. However, for atom at a temperature of 6 μK, the thermal momentum is about $2\hbar k_{eff}$. It is not obvious whether this thermal momentum is sufficiently small in comparison to the total momentum transfer for this technique to improve the performance of the PSI-LMT significantly. This issue will be investigated in the future.

## 5. Conclusion

In a point source interferometer (PSI), atoms are split and recombined by applying a temporal sequence of Raman pulses during the expansion of a cloud of cold atoms behaving approximately as a point source. The PSI can work as a sensitive multi-axes gyroscope that automatically filters out the signal from accelerations, thus making it an attractive system for practical rotation sensing. The phase shift arising from rotations is proportional to the momentum transferred to each atom from the Raman pulses. Here, we have investigated the degree of enhancement in sensitivity that could be achieved by augmenting the PSI with large momentum transfer (LMT) employing a sequence of many Raman pulses with alternating directions. Contrary to the conventional approach used for describing a PSI, we have employed a model under which the motion of the center of mass of each atom is described quantum mechanically. We have shown how increasing Doppler shifts lead to imperfections, thereby limiting the visibility of the signal fringes. We have also shown that this effect can be suppressed by increasing the effective Rabi frequencies of the Raman pulses. For a given value of the effective Rabi frequency, we show that there is an optimum value for the number of pulses employed, beyond which the net enhancement in sensitivity begins to decrease. With LMT, the total duration of the pulses can be much longer than the conventional case, making the effect of spontaneous emission highly relevant. For a given one-photon Rabi frequency, a larger detuning decreases the effective Rabi frequency, but reduces spontaneous emission. Therefore, there exists an optimal detuning dependent on the number of pulses applied. For a given value of the one-photon Rabi frequency, employing the optimal detuning, we show that there is an optimum value for the number of pulses used, beyond which the net enhancement in sensitivity begins to decrease. For a one-photon Rabi frequency of 200 MHz, for example, the peak value of the factor of enhancement in sensitivity is ~39, for a momentum transfer that is ~69 times as large as that for a conventional PSI. We also find that this peak value scales as the one-photon Rabi frequency to the power of 4/5. It is anticipated that composite pulses[18] or pulses employing adiabatic rapid passage[17] or optimal quantum control[33], which make the transfer efficiency less sensitive to detuning errors and intensity inhomogeneities, would further increase the peak enhancement in sensitivity. This will be explored in future work.

**Acknowledgment:** This work has been supported by NASA grant numbers 80NSSC19C0440 and 80NSSC20C0161, and ONR grant number N00014-19-1-2181.

**References**

[1] S.M. Dickerson et al., "Multiaxis inertial sensing with long-time point source atom interferometry," Phys. Rev. Lett., 111, 083001 (2013).
[2] G.W. Hoth et al., "Point source atom interferometry with a cloud of finite size," Applied Physics Letters, 109 (2016).
[3] A. Sugarbaker et al., "Enhanced Atom Interferometer Readout through the Application of Phase Shear," Phys. Rev. Lett. 111, 113002 (2013).
[4] C. Avinadav et al., "Rotation sensing with improved stability using point-source atom interferometry," Phys. Rev. A 102, 013326 (2020).
[5] T. Kovachy et al., "Quantum superposition at the half-metre scale," Nature, 528, 530-533 (2015).
[6] P. Asenbaum et al., :Phase Shift in an Atom Interferometer due to Spacetime Curvature across its Wave Function," Phys. Rev. Lett. 118, 183602 (2017).
[7] C.J. Bordé, "Atomic interferometry with internal state labelling," Phys. Lett. A 140, 10 (1989).
[8] M. Kasevich and S. Chu, "Atomic interferometry using stimulated Raman transitions," Phys. Rev. Lett. 67, 181 (1991).
[9] G.W. Hoth, B. Pelle, J. Kitching, and E.A. Donley, "Trade-offs in size and performance for a point source interferometer gyroscope," 2017 IEEE International Symposium on Inertial Sensors and Systems (INERTIAL), IEEE, pp. 4-7 (2017).
[10] R. Sarkar, R. Fang and S.M. Shahriar, "High-Compton-frequency, parity-independent, mesoscopic Schrödinger-cat-state atom interferometer with Heisenberg-limited sensitivity," Phys. Rev. A, 98 (2018).
[11] M.S. Shahriar et al., "Continuously Guided Atomic Interferometry Using a Single-Zone Optical Excitation: Theoretical Analysis," Optics Communications, Volume 243, Issue 1-6, p. 183-201, 21 (2004).
[12] G.M. Tino and M.A. Kasevich, "Atom interferometry," IOS Press (2014).
[13] E.T. Smith et al., "Velocity rephased longitudinal momentum coherences with differentially detuned separated oscillatory fields," Phys. Rev. Lett. 81, 1996 (1998).
[14] R. Golub and S. Lamoreaux, "Elucidation of the neutron coherence length and a matter-wave sideband interferometer," Physics Letters A, 162, 122-128 (1992).
[15] L. You and M. Holland, "Ballistic expansion of trapped thermal atoms," Phys. Rev. A, 53, R1 (1996).
[16] J. M. McGuirk, M. J. Snadden, and M. A. Kasevich, "Large Area Light-Pulse Atom Interferometry," Phys. Rev. Lett. **85**, 4498 (2000).
[17] K. Kotru, D.L. Butts, J.M. Kinast and R.E. Stoner, "Large-Area Atom Interferometry with Frequency-Swept Raman Adiabatic Passage," Phys. Rev. Lett. 115, 103001 (2015).
[18] D.L. Butts et al., "Efficient broadband Raman pulses for large-area atom interferometry," JOSA B, 30, 922-927 (2013).
[19] S.-w. Chiow, T. Kovachy, H.-C. Chien, and M. A. Kasevich, "102$\hbar$k Large Area Atom Interferometers," Phys. Rev. Lett. **107**, 130403 (2011).
[20] C. Overstreet, P. Asenbaum, T. Kovachy, R. Notermans, J. M. Hogan, and M. A. Kasevich, "Effective Inertial Frame in an Atom Interferometric Test of the Equivalence Principle," Phys. Rev. Lett. **120**, 183604 (2018).
[21] P. Cladé, S. Guellati-Khélifa, F. Nez, and F. Biraben, "Large Momentum Beam Splitter Using Bloch Oscillations," Phys. Rev. Lett. **102**, 240402 (2009).
[22] H. Müller, S.-w. Chiow, S. Herrmann, and S. Chu, "Atom Interferometers with Scalable Enclosed Area," Phys. Rev. Lett. **102**, 240403 (2009).
[23] G. D. McDonald et al., "80$\hbar$k momentum separation with Bloch oscillations in an optically guided atom interferometer," Phys. Rev. Lett. **88**, 053620 (2013).
[24] M. Gebbe et al., "Twin-lattice atom interferometry," arXiv:1907.08416 (2019).
[25] H. Müller, S.-w. Chiow, Q. Long, S. Herrmann, and S. Chu, "Atom Interferometry with up to 24-Photon-Momentum-Transfer Beam Splitters," Phys. Rev. Lett **100**, 180405 (2008).
[26] K. Kotru, D. L. Butts, J. M. Kinast, and R. E. Stoner, "Large-Area Atom Interferometry with Frequency-Swept Raman Adiabatic Passage," Phys. Rev. Lett. 115, 103001 (2015).
[27] P. Hemmer, M.S. Shahriar, V. Natoli, and S. Ezekiel, "AC-Stark Shifts in a Two Zone Raman interaction," J. of the Opt. Soc. of Am. B, 6, 1519 (1989).
[28] J. Rudolph et al., "Large Momentum Transfer Clock Atom Interferometry on the 689 nm Intercombination Line of Strontium," Phys. Rev. Lett. 124, 083604 (2020)
[29] M.S. Shahriar et al., "Dark-State-Based Three-element Vector Model for the Resonant Raman Interaction," Phys. Rev. A. 55, 2272 (1997).
[30] Y.-J. Chen et al., "Single-Source Multiaxis Cold-Atom Interferometer in a Centimeter-Scale Cell," Physical Review Applied, 12, 014019 (2019).

[31] D. Steck, 'Rubidium 87 D Line Data," https://steck.us/alkalidata/rubidium87numbers.1.6.pdf

[32] Such a configuration would produce differences in the light shifts experienced by the two ground states. This can be compensated for by adjusting the two-photon detuning of the Raman beams. Alternatively, the relative intensities on the two legs can be adjusted to balance the light shifts.

[33] J. C. Saywell et al., "Optimal control of mirror pulses for cold-atom interferometry," Phys. Rev. A 98, 023625 (2018).